\documentclass[12pt]{article}
\makeatletter \@addtoreset{equation}{section} \makeatother
\renewcommand{\theequation}{\thesection.\arabic{equation}}
\newcommand{\ba}{\begin{array}}
\newcommand{\ea}{\end{array}}
\newcommand{\beq}{\begin{equation}}
\newcommand{\eeq}{\end{equation}}
\newcommand{\bea}{\begin{eqnarray}}
\newcommand{\eea}{\end{eqnarray}}

\def\bce{\begin{center}}
\def\ece{\end{center}}

\def\nonu{\nonumber}

\def\pa{\partial}

\def\be{\beta}

\def\la{\lambda}

\def\eps6{{\displaystyle \mathop{\epsilon}^{6}}{}}
\def\g6{{\displaystyle \mathop{g}^{6}}{}}

\def\nab6{{\displaystyle \mathop{\nabla}^{6}}{}}

\def\0{{\sst{(0)}}}
\def\1{{\sst{(1)}}}
\def\2{{\sst{(2)}}}
\def\3{{\sst{(3)}}}
\def\4{{\sst{(4)}}}
\def\5{{\sst{(5)}}}
\def\6{{\sst{(6)}}}
\def\7{{\sst{(7)}}}
\def\8{{\sst{(8)}}}

\def\ba{\begin{array}}
\def\ea{\end{array}}
\def\beq{\begin{equation}}
\def\eeq{\end{equation}}
\def\be{\begin{equation}}
\def\ee{\end{equation}}

\def\la{\lambda}
\def\eps{\epsilon}

\def\ba{\begin{array}}
\def\ea{\end{array}}
\def\beq{\begin{equation}}
\def\eeq{\end{equation}}
\def\be{\begin{equation}}
\def\ee{\end{equation}}

\def\la{\lambda}
\def\eps{\epsilon}

\def\eps6{{\displaystyle \mathop{\epsilon}^{6}}{}}

\def\nab6{{\displaystyle \mathop{\nabla}^{6}}{}}

\newcommand{\bean}{\begin{eqnarray*}}
\newcommand{\eean}{\end{eqnarray*}}

\begin{document}
\thispagestyle{empty} \addtocounter{page}{-1}
   \begin{flushright}
\end{flushright}

\vspace*{1.3cm}
  
\centerline{ \Large \bf   
The OPEs of Spin-4 Casimir Currents } 
\centerline{ \Large \bf 
in the  Holographic $SO(N)$ Coset Minimal Models}
\vspace*{1.5cm}
\centerline{{\bf Changhyun Ahn } and {\bf Jinsub Paeng}
} 
\vspace*{1.0cm} 
\centerline{\it 
Department of Physics, Kyungpook National University, Taegu
702-701, Korea} 
\vspace*{0.8cm} 
\centerline{\tt ahn@knu.ac.kr, \qquad jdp2r@knu.ac.kr 
} 
\vskip2cm

\centerline{\bf Abstract}
\vspace*{0.5cm}

We compute the operator product expansion (OPE) 
between the spin-$4$ current and itself in the $WD_4$ coset minimal model
with $SO(8)$ current algebra. 
The right hand side of this OPE contains the spin-$6$ Casimir 
current which is also a generator of $WD_4$ coset minimal model. 
Based on this $N=8$ result,  
we generalize the above OPE for the general $N$(in the 
$WD_{\frac{N}{2}}$ coset minimal model) by using two $N$-generalized 
coupling constants initiated by 
Hornfeck sometime ago: 
the simplest OPE for the lowest higher spin currents.
We also analyze the similar OPE in the $WB_3$(and $WB_{\frac{N-1}{2}}$) 
coset minimal model with $SO(7)$ current algebra.
The large $N$ 't Hooft limits are discussed.
Our results in two dimensional conformal field theory 
provide the asymptotic symmetry, at the quantum level, 
of the higher spin 
$AdS_3$ gravity found by Chen et al.

\baselineskip=18pt
\newpage
\renewcommand{\theequation}
{\arabic{section}\mbox{.}\arabic{equation}}

\section{Introduction}

In the Gaberdiel and Gopakumar proposal \cite{GG,GG1},
the $W A_{N-1}$ minimal model conformal field theory is dual, in the 't Hooft $\frac{1}{N}$ expansion,
to the higher spin theory of Vasiliev on the $AdS_3$ coupled to one complex scalar.
The higher spin gauge fields in the bulk $AdS_3$ couple to a conserved higher spin currents 
(whose charges form an extended global symmetry of the conformal field theory)
in the boundary theory. See the recent review papers \cite{GG2,AGKP}. 

The $SU(N)$ spin-$3$ Casimir construction
in the $W A_{N-1}$ minimal model (described in terms of a coset) has been found in \cite{BBSS2}.
The four independent cubic terms  are made of 
the spin-$1$ currents in the two factors in the numerator of 
the coset.
Then the OPE between the spin-$3$ current and itself 
generates the spin-$4$ current \cite{Ahn2011} which consists of 
quartic terms and quadratic terms with derivatives in terms of above spin-$1$ currents.
Furthermore, 
the $SO(N)$ spin-$4$ Casimir construction in the $WD_{\frac{N}{2}}$ and $W B_{\frac{N-1}{2}}$ minimal models \cite{LF}
is found in \cite{Ahn1202} with the observation of \cite{Ahn1106,GV}. 
The quartic terms, cubic terms with one derivative and quadratic terms with 
two derivatives in the spin-$4$ current (with two unknown coefficient functions) 
are given in terms of the spin-$1$ currents in the two factors in the numerator
of the coset. 
It is natural to ask what happens when one computes the OPE between the spin-$4$ current and itself.

In this paper, 
we compute the OPE 
between the spin-$4$ Casimir current and itself, the simplest OPE for the lowest higher spin currents,  
in the $WD_{\frac{N}{2}}$ coset minimal model
with $N=8$. 
During this computation, the spin-$6$ Casimir current arises on the right hand side of this OPE
and the two unknown coefficients in the spin-$4$ current are fixed. 
We would like to generalize the above OPE for the general $N$ by using two $N$-generalization 
coupling constants \cite{Hornfeck}.
We also analyze the similar OPE in the $WB_{\frac{N-1}{2}}$ 
coset minimal model.

As described in \cite{Hornfeck}, 
the findings over there  did not answer for what kinds of 
the field contents for $W D_4$ (or $W B_3$) 
algebra are present. 
In this paper, one sees the field contents for the higher spin currents,
explicitly, living in the specific coset model we are considering.
In \cite{CGKV}, they started with the most general ansatz for the OPEs between 
the spin-$4$, spin-$6$, spin-$8$ currents and determined the various 
structure constants using the Jacobi identities between these higher spin 
currents. This can be classified by 
the approach $1$ in the context of \cite{BS}. 
With the identification of 
the above algebra as Drinfeld-Sokolov reduction (can be described as
the approach $2$ in \cite{BS}) 
of higher spin algebra, they applied to the coset model  by comparing 
the central charge and the self-coupling constant of the spin-$4$ current.  
However, it is not clear how one can see the explicit form for the higher spin
currents on the coset model.
Since the approach $3$ in \cite{BS} we are using
is based on the specific model and the higher spin 
currents are made of the fields in the coset model, one 
can analyze the zero mode eigenvalue equation which is necessary to 
describe 
the three-point function with real scalar as in \cite{GH,GGHR,Ahn1202}.
Note that the zero modes satisfy the commutation relations of the 
underlying finite dimensional Lie algebra.

According to the result of \cite{BBSS1,BS},
one expects 
that the additional currents as well as $W A_{N-1}$ (or $W D_{\frac{N}{2}}$) 
currents with arbitrary levels appear.
As one puts one of the levels
as one, then this extra current in the OPE disappears completely. 
In the spirit of \cite{GG2,Ahn1211}, 
one can think of more general algebra rather than
conventional $W A_{N-1}$ (or $W D_{\frac{N}{2}}$) 
algebra: the existence of additional higher spin currents.
One of the levels in this general coset model is not equal to $1$.
Then one cannot use the isomorphism between the coset construction and the 
Drinfeld-Sokolov reduction, explained in \cite{CGKV}, 
because this isomorphism restricts 
to have one of the levels as $1$ \cite{BS}. 
This implies that for this general coset model with levels $(k,l)$, 
one cannot follow the 
procedure developed in \cite{CGKV} directly
\footnote{
In other words, the direct construction using the Jacobi identity 
(without knowing the realization of the higher spin currents) is not 
connected to the Casimir (coset) constuction directly.  
This implies that even though one has the self-coupling constant for the 
spin-$4$ current from the work of \cite{Hornfeck,CGKV},
the coset construction itself (a realization in the specific coset model) 
is interesting in order 
to identify a complete set of 
generating currents which will have larger symmetry.
Once we determine the lower higher spin currents using the coset construction 
explicitly, then the next undetermined 
higher spin currents can be generated, in principle, by computing 
the OPEs between the known higher spin currents repeatedly 
(i.e., by analyzing the 
various singular terms). In general, this procedure will be quite involved.
Contrary to the approach $1$ (which can be done only if one knows 
the number of currents with given spins), 
the resulting extended algebra via the approach $3$ is associative 
by construction and therefore we do not have to check the Jacobi 
identities separately. 
It would be interesting to study this more general 
coset model and its $AdS_3$ gravity dual which will be beyond the scope of 
this paper. }.
Our presentation in this paper will give some hints in order to describe 
the general coset model with levels $(k,l)$ in two dimensional CFT. 

In section 2, we review the diagonal coset minimal models and describe the spin-$2$ current with the central charge.

In section 3, by taking the OPE between the spin-$4$ currents, we construct the spin-$6$ current in the 
$W D_{\frac{N}{2}}$ minimal model with $N=8$. By realizing the presence of two structure constants which depend on 
the general $N$, the OPE between the spin-$4$ currents which is valid for any $N$ is given.  

In section 4, we describe the similar OPE in the $W B_{\frac{N-1}{2}}$ minimal model.

In section 5, we summarize what we have found and comment on future directions.

In the Appendices, for convenience,
we present the relative coefficient functions in terms of two undetermined ones for each 
minimal model. These appeared in \cite{Ahn1202} previously. 

During this preparation, we have noticed that the work of \cite{CGKV} has some overlaps with our work
although they did not consider the explicit realizations we are using.
Furthermore, we also have noticed the presence of
 a previous work by Hornfeck \cite{Hornfeck} independently before 
their paper came out. 

We are heavily using the package by Thielemans \cite{Thielemans}. 

\section{The GKO coset construction: Review}

For the diagonal coset model
\bea
\frac{G}{H} = \frac{\widehat{SO}(N)_k \oplus \widehat{SO}(N)_1}{\widehat{SO}(N)_{k+1}},
\label{coset}
\eea
the spin-$1$ fields, $J^{ab}(z)$ with level $1$ and $K^{ab}(z)$ with level $k$,
generate the affine Lie algebra $G$.
The indices $a, b= 1, 2, \cdots, N$ 
are in the representation of finite dimensional Lie algebra $SO(N)$.
Their OPEs \cite{Watts} 
are 
\bea
&& J^{ab} (z) J^{cd} (w)  =  -\frac{1}{(z-w)^2} \, (-\delta^{bc}
\, \delta^{ad}+
\delta^{ac} \, \delta^{bd}) 
\nonu \\ 
&& +  \frac{1}{(z-w)} \, 
  \left[  \delta^{bc} \, J^{ad}(w) +\delta^{ad} \, J^{bc}(w)-\delta^{ac}
  \, J^{bd}(w)-
\delta^{bd} \, J^{ac}(w) \right] + \cdots,
\label{JJcoset}
\eea
and 
\bea
&& K^{ab} (z) K^{cd} (w)  =  -\frac{1}{(z-w)^2} \, k (-\delta^{bc}
\, \delta^{ad}+
\delta^{ac} \, \delta^{bd}) 
\nonu \\ 
&& +  \frac{1}{(z-w)} \,
  \left[  \delta^{bc} \, K^{ad}(w) +\delta^{ad} \, K^{bc}(w)-\delta^{ac}
  \, K^{bd}(w)-
\delta^{bd} \, K^{ac}(w) \right] + \cdots.
\label{KKcoset}
\eea
The spin-$1$ field, $J'^{ab}(z)$ with level $(k+1)$, that generates 
the affine Lie subalgebra $H$, is given by
\bea
J'^{ab} (z) = J^{ab}(z) +K^{ab}(z).
\label{diag}
\eea

The Sugawara stress energy tensor for the coset (\ref{coset}) is, with (\ref{diag}),
\bea
T(z) & = &   -\frac{1}{4(N-1)} \, (J^{ab} J^{ab}) (z) 
 -\frac{1}{4(k+N-2)} \, (K^{ab} K^{ab}) (z) \nonu \\
& + &
\frac{1}{4(k +N-1)} \, (J'^{ab} J'^{ab}) (z).
\label{Ttilde}
\eea
The OPE between the spin-$2$ currents (\ref{Ttilde}) is given by
\bea 
T (z)\; T(w) 
& = & \frac{1}{(z-w)^4} \, \frac{c}{2} +\frac{1}{(z-w)^2} \, 2 
T(w) +
\frac{1}{(z-w)} \, \pa T(w) +\cdots.
\label{cosettt}
\eea 
The central charge in the highest singular term in (\ref{cosettt}) is given by
\bea
c  =   \frac{N}{2} \left[
1-\frac{(N-2)(N-1)}{(N+k-2)(N+k-1)} 
\right] \leq \frac{N}{2}.
\label{ctilde}
\eea
The higher spin Casimir currents of spin $4$ in $W D_{\frac{N}{2}}$ and 
$W B_{\frac{N-1}{2}}$ minimal models were 
constructed in \cite{Ahn1202}. In next sections, we will compute the OPEs
between these spin-$4$ currents in each minimal model.

\section{ The OPE between the spin-$4$ current and itself in $WD_{\frac{N}{2}}$ minimal 
model with even $N$}

In \cite{Ahn1202}, the spin-$4$ current is determined, with two unknown coefficient functions,  by 
\bea
&& 
V(z)= c_3 \, J^{cd} J^{ef} K^{cd} K^{ef}(z)+
c_8 \, J^{cd} J^{ef} K^{ce} K^{df}(z)+ c_9 \, J^{cd} K^{ef} K^{ce} K^{df}(z)
\nonu \\ 
&& +
c_{10} \, K^{cd} K^{ef} K^{ce} K^{df}(z)
 +
c_{11} \, J^{cd} J^{cd} J^{ef} J^{ef}(z) + c_{12} \, J^{cd} J^{cd} J^{ef} K^{ef}(z)
 + c_{13} \, J^{cd} J^{cd} K^{ef} K^{ef}(z) \nonu \\
&& + 
 c_{14} \, J^{cd} K^{cd} K^{ef} K^{ef}(z) 
 +
c_{15} \, K^{cd} K^{cd} K^{ef} K^{ef}(z) +
c_{18} \, J^{cd} J^{ce} K^{ef} K^{df}(z) + c_{21} \, J^{cd} J^{ce} J^{df}
J^{ef}(z) \nonu \\
&& 
+ c_{22} \, J^{cd} J^{ce} J^{df}
K^{ef}(z) 
+ d_1 \, \pa J^{ab} \pa J^{ab}(z)  
+ d_2 \, \pa^2 J^{ab}  J^{ab}(z) 
+ d_3 \, \pa K^{ab} \pa K^{ab}(z)  
+ d_4 \, \pa^2 K^{ab}  K^{ab}(z) 
\nonu \\
&& + d_5
\, \pa^2 J^{ab}  K^{ab}(z) 
+ d_6 \, \pa J^{ab} \pa K^{ab}(z) 
+
d_7 \, J^{ab} \pa^2 K^{ab}(z) 
+ d_8 \, 
J^{ab} \pa J^{ac} K^{bc}(z) 
\nonu \\
&& + d_9 \,
J^{ab} K^{ac} \pa K^{bc}(z),
\label{VD}
\eea
where the coefficient functions are given in (\ref{coeffwd}).
Let us first consider the particular $N=8$ case in the $W D_{\frac{N}{2}}$ 
minimal model 
\footnote{\label{wd4} 
In this case, the currents are given by 
the spin-$4$, the spin-$6$ currents as well as the spin-$2$ current. Moreover, there exists one additional 
spin-$4$ current. For general $N$, this last current has a spin-$\frac{N}{2}$. This implies that for 
large $N$ behavior, the OPEs between the lower higher spins in the $WD_{\frac{N}{2}}$ minimal model 
do not contain this spin-$\frac{N}{2}$ field. However, for 
$W D_4$ minimal model, we will see that the extra spin-$4$ current does not 
appear in the OPE between the other spin-$4$ currents. 
This can be explained by the existence of 
the outer ${\bf Z}_2$ automorphism \cite{Honecker1,CGKV}
under which the spin-$2$, spin-$4$ and spin-$6$ currents are even while 
the extra spin-$4$ current is odd. For general $N$, the 'orbifold' 
subalgebra of $W D_{\frac{N}{2}}$ \cite{BEHHH} is generated by 
the quadratic term in the extra spin-$\frac{N}{2}$  (of spin $N$) 
as well as its higher derivative terms,
in addition to the above spin-$2$, spin-$4$, $\cdots$, spin-$(N-2)$.}.
It is straightforward to compute 
the OPE between this spin-$4$ current (\ref{VD}) and itself for
given the basic OPEs from (\ref{JJcoset}) and (\ref{KKcoset}).
We present the final result and then describe the details
we use
\bea
&& V(z) \; V(w) =\frac{1}{(z-w)^8} \, \frac{c}{4} +\frac{1}{(z-w)^6}
\, 2 \,T(w) +\frac{1}{(z-w)^5} \, \frac{1}{2} \, 2 \, \pa T(w) 
\nonu \\
&& +\frac{1}{(z-w)^4} \, \left[ \frac{3}{20} \, 2 \, \pa^2 T + 
\frac{42}{(5c+22)} \, \left( T^2 -\frac{3}{10} \pa^2 T \right) 
+ C_{44}^{4} \, V
\right] (w) \nonu \\
&& +\frac{1}{(z-w)^3} \, \left[ \frac{1}{30} \, 2 \, \pa^3 T + 
\frac{1}{2} \, \frac{42}{(5c+22)} \, \pa 
\left( T^2 -\frac{3}{10} \pa^2 T \right) 
+ \frac{1}{2}  \, C_{44}^{4} \, \pa V  \right] (w) \nonu \\
&&  +\frac{1}{(z-w)^2} \, \left[ \frac{1}{168} \, 2 \, \pa^4 T + 
\frac{5}{36} \, 
\frac{42}{(5c+22)} \, \pa^2 \left( T^2 -\frac{3}{10} \pa^2 T \right) 
+\frac{5}{36}  \, C_{44}^{4}\, \pa^2 V  \right. \nonu \\
&&  + \frac{24(72c+13)}{(5c+22)(2c-1)(7c+68)} \, \left( T (T^2 -\frac{3}{10} 
\pa^2 T) -\frac{3}{5} \pa^2 T T +\frac{1}{70} \pa^4 T \right) \nonu \\
&& -\frac{(95c^2+1254c-10904)}{6(5c+22)(2c-1)(7c+68)} \, \left( \frac{1}{2}
\pa^2 (T^2 -\frac{3}{10} \pa^2 T) -\frac{9}{5} \pa^2 T T +\frac{3}{70} 
\pa^4 T  \right) \nonu \\
&& \left. + \frac{28}{3(c+24)}  \, C_{44}^{4} \, \left( T V - \frac{1}{6} \pa^2 V \right)
 + C_{44}^{6} \, W \right] (w) \nonu \\
&&  +\frac{1}{(z-w)} \, \left[ \frac{1}{1120} \, 2 \, \pa^5 T + 
\frac{1}{36} \, 
\frac{42}{(5c+22)} \, \pa^3 \left( T^2 -\frac{3}{10} \pa^2 T \right) 
+\frac{1}{36}  \, C_{44}^{4} \, \pa^3 V  \right. \nonu \\
&&  + \frac{1}{2} \, \frac{24(72c+13)}{(5c+22)(2c-1)(7c+68)}  \, \pa
\left( T (T^2 -\frac{3}{10} 
\pa^2 T) -\frac{3}{5} \pa^2 T T +\frac{1}{70} \pa^4 T \right) \nonu \\
&& -\frac{1}{2} \, 
\frac{(95c^2+1254c-10904)}{6(5c+22)(2c-1)(7c+68)}  \, \pa 
\left( \frac{1}{2}
\pa^2 (T^2 -\frac{3}{10} \pa^2 T) -\frac{9}{5} \pa^2 T T +\frac{3}{70} 
\pa^4 T  \right) \nonu \\
&& \left. + \frac{1}{2} \, \frac{28}{3(c+24)}  \, C_{44}^{4}\,  \pa 
\left( T V - \frac{1}{6} \pa^2 V \right) + \frac{1}{2} \, C_{44}^{6} \, \pa W
 \right] (w)
+ \cdots.
\label{ansWD}
\eea
Let us first consider the highest singular term.
From the explicit structure of the eighth-order pole, one obtains
the following expression
\bea
192 k (2 + k) (4 + k) (6 + k) (7 + k) (9 + k) 
(11 + k) (13 + k) \, c_{10}^2.
\label{equation-order8}
\eea
By normalizing 
that this expression (\ref{equation-order8}) is  equal to $\frac{c}{4}$
with 
\bea
c_{N=8} =\frac{4 k (13 + k)}{(6 + k) (7 + k)}
\label{cn8}
\eea
coming from the central charge (\ref{ctilde}), one 
determines the unknown coefficient $c_{10}$ as follows:
\bea
c_{10}^2 = \frac{1}{192 (2+k) (4+k) (6+k)^2 (7+k)^2 (9+k) (11+k)}=
\frac{(-4+c)^4}{130056192 (24+c) (22+5 c)}.
\label{c10}
\eea
We also express the coefficient $c_{10}$ 
in terms of the central charge (\ref{cn8}) for convenience.
Therefore, the central term is given by $\frac{c}{4}$ as in (\ref{ansWD}).

The next singular term, seventh pole, does not have any nonzero fields.
In the sixth order pole, 
there exists a spin-$2$ current $T(w)$ 
with coefficient $2$ which is a structure constant between 
the spin-$4$, the spin-$4$ and the spin-$2$ currents. 
In the fifth order pole, one obtains the descendant field $\pa T(w)$
of $T(w)$ with the coefficient function $\frac{1}{2} \times 2 =1$
where the relative coefficient function $\frac{1}{2}$ 
is known from the spins of the current $V(z)$, the current $V(w)$, the current
$T(w)$ and the number of derivatives in the descendant field 
\cite{Bowcock,BFKNRV,BS,Ahn1211}.

In the fourth order pole, 
one also has a descendant field originating from the spin-$2$ current $T(w)$ 
appeared in the higher-order singular term with known coefficient $\frac{3}{20} \times 2 =\frac{3}{10}$. 
Furthermore, one expects that there exist a spin-$4$ quasi primary field as well as the spin-$4$ primary 
field $V(w)$ by remembering the OPE of spin-$4$ currents in the extended conformal algebra 
\cite{Bouwknegt,HT,Zhang,BFKNRV,KW}, 
denoted by
${\cal W}(2,4)$ along the line of \cite{BS}, where the 
higher spin current is of  spin-$4$ \footnote{If one does not know the field contents in the 
fourth-order pole, one can follow the method done in \cite{Ahn1211}. One performs the OPE between the 
spin-$2$ current $T(z)$  
and the fourth-order pole subtracted by $\frac{3}{10} \pa^2 T(w)$ and focus on the 
fourth-order pole. Then one has nonzero $T(w)$ term on the right hand side of
OPE. This indicates that one can consider the extra 
quasi-primary field containing $T^2(w)$ term with derivative term because the 
OPE between $T(z)$ and $T^2(w)$ provides a term $T(w)$ in the fourth order 
pole. 
The coefficient $-\frac{3}{10}$ in the derivative term can be checked via 
the vanishing of third-order pole in the OPE with $T(z)$. Then it is simple to compute the OPE between 
$T(z)$ and  the fourth-order pole subtracted by $\frac{3}{10} \pa^2 T(w)+ c_1 
(T^2 -\frac{3}{10} \pa^2 T)(w)$. Once again, the primary field 
condition fixes the constant as $c_1=
\frac{42}{(5c+22)}$. Of course, one should compute the OPE between 
$T(z)$ and $\pa^2 T(w)$ explicitly in order to obtain this result. }. 
Then it turns out that the fourth-order pole is given by the one in (\ref{ansWD}) 
and the self-coupling constant for the spin-$4$
is given by
\bea
(C_{44}^{4})^2 = \frac{12 (4+k) (9+k)}
{(2+k) (11+k)} =\frac{18 (c+24)}{(5 c+22)},
\label{c444}
\eea
where we also write down $C_{44}^{4}$ in terms of the central charge.
One can also determine the remaining unknown coefficient function $c_8$ as follows:
\bea
c_8 =
\frac{(6+k)}{105} \left[ (7+k) \sqrt{(4+k) (9+k) (148+k (13+k))}+
 (4+k) (133+k (16+k))\right] c_{10}
\label{c8}
\eea
with (\ref{c10}).
Of course, the structure constant (\ref{c444}) is different from the 
corresponding one in 
the ${\cal W}(2,4)$ algebra as in \cite{BS}.
For the $W D_4$ algebra, the field contents are given by 
the spin-$4$, the spin-$4$ and the spin-$6$ currents as in footnote \ref{wd4}
while ${\cal W}(2,4)$ algebra contains only the spin-$4$ current.
The above structure constant (\ref{c444}) 
coincides with the general-$N$ dependent structure constant 
found by Hornfeck by substituting $N=8$ with  
\bea
(C_{44}^{4})^2 & = & \frac{n}{d}, \nonu \\
n & \equiv &  18 \left[2 c^2 (-18+(-2+N) N)+2 N (-28+N (5+6 N)) \right. \nonu \\ 
& + & \left. 3 c 
(-8+N (80+N (-49+6 N)))\right]^2, \nonu \\ 
d & \equiv & (22+5 c) (c+(-5+c) N+4 N^2 ) 
(c (-4+N) (-3+N)+N (-5+2 N)) \nonu \\
& \times & (2 c (2+N)+(-4+N) (-2+3 N)).
\label{c444final}
\eea
Recently, this expression is reproduced in \cite{CGKV}.

In the third-order pole, there are no additional spin-$5$ quasi-primary fields.
There are one descendant field coming from $T(w)$ and two descendant fields 
coming from spin-$4$ quasi primary and primary fields.

Now we are 
ready to consider the next second-order singular term \footnote{It took several months to 
compute 
the complete pole structures (up to second-order pole) with several personal 
computers. Although we have not 
checked the first-order pole explicitly, we expect that the first-order pole in (\ref{ansWD}) is correct.
The point is whether there exists an extra quasi-primary field of spin-$7$ or not. Since $V(z) \; V(w)=
V(w)\; V(z)$, one can reverse the arguments $z$ and $w$ in (\ref{ansWD}) with the insertion 
of some quasi-primary field of spin-$7$ and use Taylor expansions about 
the coordinate $w$. Then we have an explicit 
expression as in the Appendix of 
\cite{BBSS1}. All the higher order 
terms greater than order-$1$ can appear as the derivative 
terms at the first-order pole. It turns out that the first-order term in 
$V(w) \; V(z)$ appears 
as the first-order pole term 
in (\ref{ansWD}) and above spin-$7$ field with 
opposite sign. 
Therefore, one realizes that by comparing both sides, 
the above quasi-primary field of spin-$7$ 
vanishes.     }.
One expects that there should be a spin-$6$ primary field which is a generator of 
$WD_4$ minimal model.
The first line of the second-order pole in (\ref{ansWD})
is known and the remaining three terms are characterized by three spin-$6$ quasi-primary
fields 
\footnote{In \cite{BS}, the expression for $\Omega_{BS}$ in $(5.11)$ is not right.
The correct one is 
$\Omega_{BS}(w) = 
T(T^2-\frac{3}{10} \pa^2 T)(w) -\frac{3}{5} \pa^2 T T(w) +\frac{1}{70} \pa^4 T(w)$ as in 
(\ref{ansWD}). Also 
note that the notation for the normal ordering we are using here is different from the one in \cite{BFKNRV} as emphasized in \cite{Ahn1211}. Sometimes there are several ways to express the 
quasi-primary field by using the identities $T \pa^2 T(z) = \pa^2 T T(z) +\frac{1}{6} \pa^4 T(z)$ and 
$\pa^2 (T^2 -\frac{3}{10} \pa^2 T)(z) = 2 \pa T \pa T(z) + 2 \pa^2 T T(z) -\frac{2}{15} \pa^4 T(z)$. 
That is, $H_{BS}(z)$ corresponds to $\Omega(z)$ of \cite{BFKNRV}, $P_{BS}(z)$ corresponds to $-\frac{5}{9} 
\Gamma(z)$, and $(\Omega_{BS}-\frac{1}{3} P_{BS})(z)$ corresponds to $\Delta(z)$. Note that any linear 
combinations of quasi-primary fields for given spin 
provide a different quasi-primary field. The convention for the quasi-primary
fields in \cite{Zhang} is the same as the one in \cite{BS} while the 
convention for the same quantity in \cite{BFKNRV} is the same as 
those in \cite{CGKV,HT}. }. 
Although the $V$-independent terms are characterized by
$\pa^4 T(w), \pa^2 T T(w), T^3(w) $ and $\pa T \pa T(w)$(these are 
all possible spin-$6$ fields coming from
th spin-$2$ current $T(w)$), it is very important to 
split two descendant terms and two quasi-primary fields in order to find out the new quasi-primary 
fields for given pole.
These also arise in the ${\cal W}(2,4)$ extended conformal algebra.   
Then finally, we are left with a spin-$6$ primary field 
where
\bea
C_{44}^{6} \, W(z) = \mbox{coeff}(k) \,
J^{12} J^{12} J^{12} J^{12} J^{12} J^{12}(z) + \cdots,
\label{spin6}
\eea
and $\mbox{coeff}(k)$ is  a complicated function of $k$ and we do not 
present it here.

In order to determine the normalization for the spin-$6$ current, one should compute the highest singular 
term from the OPE between the current (\ref{spin6}) and itself.
However, it is not possible, at the moment, 
to do this because the spin-$6$ current has too 
many terms.
By demanding that this central term should be equal to $\frac{c}{6}$, 
one expects that one should obtain the normalization 
factor as follows:
\bea
(C_{44}^{6})^2 =\frac{6 (-1+k) (14+k) (24+5 k) (41+5 k)}{(-6+k (13+k)) (119+4 k (13+k))}  
=\frac{12 (c-1) (11 c+656)}{(2 c-1) (7 c+68)}. 
\label{coeff446}
\eea
This structure constant (\ref{coeff446}) is taken from 
the more general $N$ dependent 
expression found by Hornfeck
\bea
(C_{44}^{6})^2 & = & \frac{n}{d}, \nonu \\
n & \equiv & 12 (-1+c) (22+5 c)^2 (c (-6+N) (-5+N)+2 N (-7+2 N)) \nonu \\
& \times & (2 c (4+N)+(-8+3 N) (-4+5 N)) 
(c (3+N)+2 N (-7+6 N)), \nonu \\ 
d & \equiv & (24+c) (-1+2 c) (68+7 c) (c+(-5+c) N+4 N^2) \label{c446final} \\
& \times & (c (-4+N) (-3+N)+N (-5+2 N)) 
(2 c (2+N)+(-4+N) (-2+3 N)). 
\nonu
\eea
As pointed out by Hornfeck \cite{Hornfeck}, there exists also an extended conformal algebra 
${\cal W}(2,4,6)$ in \cite{KW} which is nothing to do with the present case but is related to the next 
example $WB_1$ coset minimal model. By substituting $N=3$ into the formula 
(\ref{c444final}) and (\ref{c446final}), the author 
could obtain the structure constants in \cite{KW} exactly. 
Note that the spin-$4$ and spin-$6$ currents are made of 
stress energy tensor and its ${\cal N}=1$ superpartner of spin $\frac{3}{2}$ 
\cite{BS}. The explicit form is given in \cite{Honecker}.
Furthermore, the author of \cite{Hornfeck} checked that 
the classical $c \rightarrow \infty$ limit of (\ref{c444final})
coincides with the one in \cite{FRS} where the structure constant 
can be obtained from the one from $W A_{N-2}$ minimal model \footnote{
\label{c444limit}
The 
self-coupling structure constant between the spin-$4$ currents 
in the $W A_{N-2}$ minimal model is given by \cite{Hornfeck1992,GG1} $
(C_{44}^{4})^2 =
\frac{36 \left(-24-48 c-18 c^2+224 N+204 c N-2 c^2 N-130 N^2-129 c N^2+c^2 N^2+12 N^3+18 c N^3\right)^2}{(2+c) (22+5 c) (-4+N) (-3+N) \left(2+c-7 N+c N+3 N^2\right) \left(4+2 c-18 N+c N+8 N^2\right)}$. By taking the $c \rightarrow \infty$
limit, this leads to $\frac{36 \left(-18-2 N+N^2\right)^2}{5 (-4+N) 
(-3+N) (1+N) (2+N)}$ which is equal to the corresponding limit of 
(\ref{c444final}) \cite{FRS}.}.
Note that there exists a factor $(c-1)$ in the structure constant in 
(\ref{c446final}). In other words, $c=1$ implies that $k=1$ or $k=2(1-N)$.
Then, for $k=1$, the structure constant $C_{44}^{6}$ vanishes.
This is kind of `minimal' extension of conformal algebra \cite{Ahn1211} where 
the only higher spin current is of spin $4$ while 
the higher spin current of spin-$6$ vanishes. 

We claim that the lowest OPE between the spin-$4$ current in the $WD_{\frac{N}{2}}$ coset 
minimal model is characterized by (\ref{ansWD}) where the central charge is given by 
(\ref{ctilde}), the spin-$4$ current is given by (\ref{VD}), 
the spin-$2$ current is given by (\ref{Ttilde}), the structure constant $C_{44}^{4}$ is given by
(\ref{c444final}), and the structure constant $C_{44}^{6}$ is given by 
(\ref{c446final}). For the spin-$4$ and spin-$6$ currents, we have found 
for particular $N=8$ case in the $W D_{\frac{N}{2}}$ minimal model 
(and $N=6$ case)
\footnote{We have checked the OPE (\ref{ansWD}) when $N=6$ and realize that 
the result is exactly the same as (\ref{ansWD}) by replacing the central charge 
$c_{N=6}=\frac{3 k (9+k)}{(4+k) (5+k)}$ and the $N=6$ for the currents and structure constants.
The field contents of $W D_{\frac{N}{2}}$ minimal model are given by the spins $2, 4, \cdots, (N-2), \frac{N}{2}$.
The naive field contents for $N=6$ are given by spins $2, 3$, and $4$ in this formula.
According to the observation of \cite{LF1}, the minimum value of $N$ for the above field contents in the 
$W D_{\frac{N}{2}}$ minimal model is equal to $8$.}.
For the spin-$4$ current at general $N$, 
there exist two unknown coefficient functions as we explained before.
It would be interesting to obtain this spin-$6$ current which holds for arbitrary $N$.
This can be done by computing the OPEs between the spin-$4$ currents (\ref{VD}) by hand.

The large $(N, k)$ 't Hooft limit provides the following limiting value for the 
central charge \cite{GG,Ahn1106,GV} 
\bea
 c \rightarrow \frac{N}{2} (1-\la^2), \qquad \la \equiv \frac{N}{N+k-2}.
\label{climit}
\eea
Furthermore, the limiting values for structure constants are obtained  
and they are
\bea
(C_{44}^{4})^2 & \rightarrow &
\frac{36 \left(-19+\lambda ^2\right)^2}{5 (-3+\lambda ) (-2+\lambda ) (2+\lambda ) (3+\lambda )},
\nonu \\
(C_{44}^{6})^2 & \rightarrow &
\frac{150 (-5+\lambda ) 
(-4+\lambda ) (4+\lambda ) (5+\lambda )}{7 (-3+\lambda ) (-2+\lambda ) (2+\lambda ) (3+\lambda )}.
\label{structlimit}
\eea
Then the OPE (\ref{ansWD}) under the large $(N,k)$ limit can be obtained
by substituting (\ref{climit}) and (\ref{structlimit}) into (\ref{ansWD}).
See also the related works \cite{Ahn1206,Ahn1208} where
the large $(N,k)$ 't Hooft limit on the OPE was described.   

Recently, the asymptotic symmetry of the truncated 
higher spin gravity in the context of
black hole \cite{CLW} turns out to be 
the classical ${\cal W}(2,4)$ algebra. 
By changing the Poisson bracket into the commutators between Fourier modes,
one obtains the various commutation relations. 
Note that the Fourier mode of normal ordered product field is defined 
as the sum of product of each Fourier mode   in \cite{BBSS1}.
On the other hand, 
one can take the classical $c \rightarrow \infty$ 
limit for the OPE in (\ref{ansWD}). 
Any composite field of order $n$ can contain 
only $\frac{1}{c^{n-1}}$-behavior term. For example, the $\pa^2 T(w)$ term
appears in the fourth-order pole. The $c$-independent term 
survives while $c$-dependent term goes away because it has $\frac{1}{c}$ 
behavior. The quasi-primary field containing $T^3(w)$ term in the 
second-order pole has cubic term, quadratic term and linear term in $T(w)$.
The overall factor behaves as $\frac{1}{c^2}$ under the large $c$ limit. 
Therefore, the only cubic term can survive.
See also \cite{BW,Ahn1206} where the similar limiting procedure was done.
Eventually, one can check the classical version of (\ref{ansWD}) matches with 
the one in \cite{CLW} by turning off the structure constant $C_{44}^{6}$
which appears in front of the spin-$6$ current on the right hand side of 
OPE
\footnote{For the $W D_4$ algebra, the structure constant (\ref{coeff446})
vanishes at $c=1$ or $c=-\frac{656}{11}$.
In \cite{BFKNRV,KW}, the ${\cal W}(2,4,4)$ algebra
has been shown to be consistent with for these values 
$c=1$ and $c=-\frac{656}{11}$.
As we take $c \rightarrow \infty$ limit with fixed $N$, 
the structure constant
behaves as 
$(C_{44}^6)^2  \rightarrow
\frac{150 (-6+N) (-5+N) (3+N) (4+N)}{7 (-4+N) (-3+N) (1+N) (2+N)}$
which is the ratio of each $c^6$ term in the denominator and numerator of
$(C_{44}^6)^2$.
Therefore, for $N=6$, this structure constant vanishes. 
 One sees the behavior of the structure constant
$C_{44}^4$ in the classical limit. According to the observation of footnote
\ref{c444limit}, 
one has $(C_{44}^4)^2 \rightarrow \frac{27}{35}$ by substituting 
$N=6$ into the formula. Note that this numerical value 
$\frac{27}{35}$ is exactly the same as the one in the classical 
${\cal W}(2,4)$ algebra because $\frac{54(c+24)(c^2-172c+1296)}{
(5c+22)(2c-1)(7c+68)} \rightarrow \frac{54}{5\cdot 2\cdot 7}=\frac{27}{35}$. 
The $W D_3$ algebra reduces to the ${\cal W}(2,4)$ algebra 
\cite{CLW}. For $N=5$, the above structure constant 
$C_{44}^6$ vanishes and from the footnote \ref{wb2eb2}, 
the structure constant $(C_{44}^4)^2$ reduces to $\frac{27}{35}$.
We thank the referee for raising this issue. }. 

In other words, our OPE (\ref{ansWD}), at the quantum level, 
provides the asymptotic symmetry algebra in the bulk theory. 
The more general analysis in higher spin $AdS_3$  gravity 
corresponding to $W D_{\frac{N}{2}}$ coset minimal model 
should produce the spin-$6$ current on the right hand
side of OPE and the quantum behavior coming from the normal ordering 
in the composite fields should appear. 
Note that the OPE (\ref{ansWD}) holds for any $N$ and is an exact (and 
complete) expression.   
Each higher spin current is made of Casimir operators that are constructed from 
the WZW currents.
The spin-$4$ current has two undetermined coefficients and the complete 
form for the spin-$6$ current is not known.

\section{ The OPE between the spin-$4$ current and itself in 
$WB_{\frac{N-1}{2}}$ 
minimal 
model with odd $N$}

In this case, the spin-$1$ field is realized by $N$ free fermions 
\cite{Watts,Ahn1992,Ahn1991}
\bea
J^{ab}(z) = \psi^a \psi^b(z).
\label{twofermion}
\eea
The OPE between the fermions is given by
\bea
\psi^a(z) \psi^b(w) =\frac{1}{(z-w)} \, \delta^{ab} +\cdots.
\label{opepsipsi}
\eea
One can easily see the OPE (\ref{JJcoset}) by using (\ref{twofermion}) and (\ref{opepsipsi}).
One takes the other OPE (\ref{KKcoset}). The Sugawara stress energy tensor is given by 
(\ref{Ttilde}) with diagonal current (\ref{diag}). The OPE satisfies (\ref{cosettt}) with the central 
charge (\ref{ctilde}).

The spin-$4$ current in \cite{Ahn1202} is, with two undetermined coefficient functions,  given by
\bea
&& 
V(z)= c_3 \, J^{cd} J^{ef} K^{cd} K^{ef}(z)+
 c_9 \, J^{cd} K^{ef} K^{ce} K^{df}(z) +
c_{10} \, K^{cd} K^{ef} K^{ce} K^{df}(z) \nonu \\
&& +
c_{11} \, J^{cd} J^{cd} J^{ef} J^{ef}(z) + c_{12} \, J^{cd} J^{cd} J^{ef} K^{ef}(z)
 + c_{13} \, J^{cd} J^{cd} K^{ef} K^{ef}(z) + 
 c_{14} \, J^{cd} K^{cd} K^{ef} K^{ef}(z) \nonu \\
&& +
c_{15} \, K^{cd} K^{cd} K^{ef} K^{ef}(z) +
c_{18} \, J^{cd} J^{ce} K^{ef} K^{df}(z)  
+ d_1 \, \pa J^{ab} \pa J^{ab}(z)  
+ d_2 \, \pa^2 J^{ab}  J^{ab}(z) \nonu \\
&& + d_3 \, \pa K^{ab} \pa K^{ab}(z)  
+ d_4 \, \pa^2 K^{ab}  K^{ab}(z) 
+ d_5
\, \pa^2 J^{ab}  K^{ab}(z) 
+ d_6 \, \pa J^{ab} \pa K^{ab}(z) 
+
d_7  \, J^{ab} \pa^2 K^{ab}(z) 
\nonu \\
&& + d_8 
\, J^{ab} \pa J^{ac} K^{bc}(z) + d_9
\, J^{ab} K^{ac} \pa K^{bc}(z),
\label{VB}
\eea
where the coefficient functions are given by (\ref{exactcoeff}).

Now one can compute the OPE $V(z) \; V(w)$ and it turns out that one has 
the equation (\ref{ansWD}). The highest singular term with $N=7$ has 
\bea
\mbox{pole} \, 8  & = & \frac{n}{d},\nonu \\
n & \equiv &  
21 k \left(720 c_9^2 (-1+k) (2+k)^2 (3+2 k)^2 (10+3 k) (21+4 k) (285+31 k)^2 \right. \nonu \\
& + &  d_8^2 
(5+k) (1320+79 k (11+k)) \nonu \\
& \times & \left. (-898722+k (201615+k (578098+k (126529+92 k (107+3 k)))))\right),
\nonu \\
d & \equiv & (2+k) (5+k) (3+2 k) (285+31 k)^2 (-250+23 k (5+6 k)),
\label{pole8}
\eea
where there exist two unknown coefficients $c_9$ and $d_8$.
Normalizing this (\ref{pole8}) to be $\frac{c}{4}$ where the central charge is
\bea
c_{N=7}=\frac{7 k (11+k)}{2 (5+k) (6+k)},
\label{centraln7}
\eea
one has one relation between the coefficients.
The seventh order pole vanishes as before.
The sixth order pole can be written in terms of $2 T(w)$ where
$T(w)$ is given by (\ref{Ttilde}) with $N=7$.

Then the unknown two coefficients are determined as
\bea
c_9^2 & = &
\frac{1320+869 k+79 k^2}{24 (2+k) (6+k)^2 (9+k) (3+2 k) (19+2 k) (10+3 k) (23+3 k)}, \nonu \\
d_8^2 & = & 
\frac{(2+k) (3+2 k) (10+3 k) (285+31 k)^2}{24 (6+k)^2 (9+k) (19+2 k) (23+3 k) (1320+869 k+79 k^2)}.
\label{c9d8}
\eea
One can write down these (\ref{c9d8}) in terms of (\ref{centraln7}) but the expressions are rather 
complicated.

From the fourth-order pole, one obtains the self-coupling constant for the spin-$4$ current 
\bea
(C_{44}^{4})^2 & = & \frac{150 (7224+6677 k+2180 
k^2+286 k^3+13 k^4)^2}{(2+k) (9+k) (3+2 k) (19+2 k) (10+3 k) (23+3 k) (1320+869 k+79 k^2)}
\nonu \\
& = & \frac{2 (4214+627 c+34 c^2)^2}{(21+4 c) (22+5 c) (19+6 c) (161+8 c)}.
\label{c444wb}
\eea
This coincides with the results \cite{Ozer} from the quantum Miura transformation.
It is easy to check that one also obtains (\ref{c444wb}) from (\ref{c444final}) by putting 
$N =7$.
The classical $c \rightarrow \infty$ limit of (\ref{c444final})
coincides with the one in \cite{FRS} where the structure constant 
can be obtained from the one from $W A_{N-1}$ minimal model
\footnote{The 
self-coupling structure constant between the spin-$4$ currents 
in the $W A_{N-1}$ minimal model is given by \cite{Hornfeck1992} $
(C_{44}^{4})^2 =
\frac{36 \left(82+45 c-19 c^2-94 N^2-75 c N^2+
c^2 N^2+12 N^3+18 c N^3\right)^2}{(2+c) 
(22+5 c) (-3+N) (-2+N) \left(-2+2 c-N+c N+3 N^2\right) 
\left(-6+3 c-2 N+c N+8 N^2\right)}
$. By taking the $c \rightarrow \infty$
limit, this leads to $ \frac{36 \left(-19+N^2\right)^2}{5 (-3+N) 
(-2+N) (2+N) (3+N)}$ which is equal to the corresponding limit of 
(\ref{c444final}) with $N$ replaced by $N+1$ \cite{FRS}.}.

Also one can read off the spin-$6$ current
\bea
C_{44}^{6} \, W(z) = \mbox{coeff}(k) \psi^a \pa^5 \psi^a(z) +\cdots,
\label{secondspin6}
\eea
where $\mbox{coeff}(k)$ is a complicated function of $k$.
One also expects that one obtains the following structure constant,
after computing the OPE between the spin-$6$ current (\ref{secondspin6}) 
and itself,
\bea
(C_{44}^{6})^2 & = & \frac{n}{d} =\frac{80 (-1+c) (49+c)^2 (22+5 c)^2 (403+22 c)}{3 (24+c) (-1+2 c) 
(21+4 c) (19+6 c) (68+7 c) (161+8 c)}, \nonu \\
n & \equiv & 5 (-1+k) (4+k)^2 (7+k)^2 (12+k) (13+4 k) (31+4 k) 
(1320+869 k+79 k^2)^2,\nonu \\
d & \equiv & (2+k) (9+k) 
(3+2 k) (19+2 k) (10+3 k) (23+3 k) (-5+11 k+k^2) 
\nonu \\
& \times & (288+121 k+11 k^2) (816+407 k+37 k^2).
\label{446wb}
\eea
One sees this expression from (\ref{c446final})
by taking $N=7$. Also this structure constant 
(\ref{446wb}) appeared in \cite{Ozer}.
The lowest OPE between the spin-$4$ current and itself 
in the $WB_{\frac{N-1}{2}}$ coset 
minimal model is characterized by (\ref{ansWD}) where the central charge is given by 
(\ref{ctilde}), the spin-$4$ current is given by (\ref{VB}), 
the spin-$2$ current is given by (\ref{Ttilde}), the structure constant $C_{44}^{4}$ is given by
(\ref{c444final}), and the structure constant $C_{44}^{6}$ is given by 
(\ref{c446final}). For the spin-$4$ and spin-$6$ currents, we have found 
for particular $N=7$ case (and $N=5$ case) 
\footnote{
\label{wb2eb2}
We have also checked for the OPEs in $WB_2$ 
minimal model \cite{FST,Ahn1991} where the spin-$6$ current contains
a term $U \pa U(w)$ and $U(w)$ is a spin-$\frac{5}{2}$ current 
\cite{Ahn1202}.
The structure constants 
$(C_{44}^{4})^2=
\frac{54 \left(-490+83 c+2 c^2\right)^2}{(25+2 c)^2 (22+5 c) (13+14 c)}$
and
$(C_{44}^6)^2 =
\frac{720 (-1+c) (115+4 c) (22+5 c)^2 (49+6 c)}{(24+c) (-1+2 c) (25+2 c)^2 (68+7 c) (13+14 c)}$ in $W B_2$ minimal model 
can be obtained from (\ref{c444final}) and 
(\ref{c446final}) by plugging 
$N=5$ respectively. There is a $(c-1)$ factor in $C_{44}^6$.
}.
As described before, when $N=3$ (i.e., $W B_1$ coset minimal model),
the structure constants (\ref{c444final}) and (\ref{c446final})
produce the previous known results in \cite{KW}
\footnote{That is, $(C_{44}^{4})^2=
\frac{54 \left(-82+47 c+10 c^2\right)^2}{(21+4 c) (22+5 c) (-7+10 c)}$ and $(C_{44}^6)^2=\frac{144 (-1+c)^2 (11+c) (22+5 c)^2 (11+14 c)}{(24+c) (-1+2 c) (21+4 c) (68+7 c) (-7+10 c)}$. There is a $(c-1)$ factor in the second structure 
constant.}. 
As before, the `minimal' extension of conformal algebra arises for $k=1$ where 
the only higher spin current is of spin-$4$ while 
the higher spin current of spin-$6$ vanishes.  

\section{Conclusions and outlook }

We have found the OPEs (\ref{ansWD}) 
between the spin-$4$ current 
and itself in the $W D_{\frac{N}{2}}$ and $W B_{\frac{N-1}{2}}$ 
coset minimal models, by checking those in 
$W D_3$ and $W D_4$ minimal models (and $W B_2$ and $W B_3$) explicitly.
These are the simplest OPEs for given minimal models. 
By using the holography between the above conformal field theory and higher
spin $AdS_3$ gravity, we expect that the bulk computation, 
at the quantum level, should possess the asymptotic symmetry corresponding to 
the OPE (\ref{ansWD}).

It is an open problem to find the correct answer for the following interesting subjects. 

$\bullet$ The full expressions of 
the spin-$4$ and the spin-$6$ currents at general $N$.
So far, the spin-$4$ current is found, for general $N$, 
up to two unknown coefficients.
This can be done only after one 
should compute the OPE between the spin-$4$ current and itself by hand. 
After doing this complicated long computation, one can extract 
the spin-$4$ current and the spin-$6$ current (up to an overall factor) 
completely. 
Or one can follow the method in \cite{Ahn1202} in order to obtain the 
spin-$6$ current by imposing that the OPE with diagonal current has no singular
term and the spin-$6$ current is primary field under the stress energy tensor.
It is nontrivial to exhaust all the possible terms coming from sextic-, 
$\cdots$,
cubic- and qudratic-terms in WZW currents. 

$\bullet$ The quantum Miura transformations and the corresponding 
higher spin currents. One can also find the higher spin currents using 
the quantum Miura transformation. 
Then it is straightforward to compute the OPE between the spin-$4$ and 
the spin-$6$ and the OPE between the spin-$6$ and itself. 
The nontrivial part in this direction is to obtain the primary fields under the
stress energy tensor by combining the nonprimary fields with fixed spins.
Also it is interesting to see whether 
the other structure constants in \cite{Hornfeck} are correct or not.

$\bullet$ Any supersymmetric extensions?
So far, the supersymmetric versions of minimal model holography 
are described in the recent works  
\cite{CHR,CG,Reyetal,HP,Ahn1206,CG1,Ahn1208,Tan,CHR1,CHR2,MZ,Peng,Hikida}.
In particular, it would be interesting to see whether 
the $W B_{\frac{N-1}{2}}$ minimal model can be generalized to 
the supersymmetric extension or not.
In the coset model (\ref{coset}), one of the level is fixed by one.
What happens if this level is given by $N$ along the line of \cite{Ahn1211}? 
As pointed out in \cite{CGKV}, it is an open problem to find out other 
supersymmetric coset minimal models. 

$\bullet$
It is natural to ask whether the next higher spin-$5$ Casimir 
current in the context of \cite{Ahn2011} can be obtained 
from the OPE between the spin-$3$ current and the spin-$4$ current or not. 
It would be interesting 
to construct the spin-$5$ current explicitly. 

\vspace{.7cm}

\centerline{\bf Acknowledgments}

This work was supported by the Mid-career Researcher Program through
the National Research Foundation of Korea (NRF) grant 
funded by the Korean government (MEST) (No. 2012-045385).
We thank Korea Institute for Advanced Study (KIAS) to access
computer facility.
CA acknowledges warm hospitality from 
the School of  Liberal Arts (and Institute of Convergence Fundamental
Studies), 
Seoul National University of Science and Technology.

\newpage

\appendix

\renewcommand{\thesection}{\large \bf \mbox{Appendix~}\Alph{section}}
\renewcommand{\theequation}{\Alph{section}\mbox{.}\arabic{equation}}

\section{The coefficients in the spin-$4$ current of
$W D_{\frac{N}{2}}$ minimal model }

The explicit coefficient functions \cite{Ahn1202} in (\ref{VD}), in terms of $c_8$ and 
$c_{10}$,
are given by
\bea
c_3 & = & 
\left(c_8 \left(k^2 (44+5 N)+44 \left(2-3 N+N^2\right)+k
    \left(-132+73 N+10 N^2\right)\right) \right. \nonu \\
&&  -2 c_{10} \left(42 k^4+k^3
    (-338+163 N)+2 (-2+N)^2 \left(76-67 N+12 N^2\right)
\right. 
\nonu \\
&& \left. \left. +k^2
    \left(1000-953 N+219 N^2\right)+k \left(-1288+1826 N-835 N^2+122
      N^3\right)\right)\right)/
\nonu \\
&& 
\left(k^2 (44+5 N)+44 \left(2-3
    N+N^2\right)+k \left(-132+73 N+10 N^2\right)\right),
\nonu \\
c_9 & = & -4 c_{10} (-2+k+N),
\nonu \\
c_{11} & = & -\left(k \left(-c_8 (-1+k) \left(5-3 N+N^2\right)
    \left(k^2 (44+5 N)+44 \left(2-3 N+N^2\right)
\right. \right. \right.
\nonu \\
&& \left. \left. \left. +k \left(-132+73 N+10
        N^2\right)\right)+2 c_{10} \left(44 (-2+N)^3 \left(4-5
        N+N^2\right) \right. \right. \right. \nonu \\
&& \left. \left. \left. +6 k^5 \left(29-15 N+7 N^2\right)+k^4
      \left(-1438+1433 N-683 N^2+163 N^3\right) \right. \right. \right.
\nonu \\
&&  +2 k (-2+N)^2
      \left(648-876 N+371 N^2-92 N^3+12 N^4\right)  \nonu \\
&&  +3 k^3
      \left(1536-2249 N+1377 N^2-471 N^3+73 N^4\right)
\nonu \\
&& \left. \left. \left. +k^2
      \left(-7096+13610 N-10495 N^2+4424 N^3-1090 N^4+122
        N^5\right)\right)\right)\right)/
\nonu \\
&& \left(2 (-1+N)^2 \left(2-5 N+2
    N^2\right) \left(k^2 (44+5 N)+44 \left(2-3 N+N^2\right)
\right. \right.
\nonu \\
&& \left. \left. +k
    \left(-132+73 N+10 N^2\right)\right)\right),
\nonu \\
c_{12} & = & 
\left(2 \left(-c_8 (-1+k) \left(5-3 N+N^2\right) \left(k^2 (44+5
      N)+44 \left(2-3 N+N^2\right) \right. \right. \right. \nonu \\
&& \left. \left. \left. + k \left(-132+73 N+10
        N^2\right)\right)+2 c_{10} \left(44 (-2+N)^3 \left(4-5
        N+N^2\right) \right. \right. \right.
\nonu \\
&&  +6 k^5 \left(29-15 N+7 N^2\right)+k^4
      \left(-1438+1433 N-683 N^2+163 N^3\right)
\nonu \\
&&  +2 k (-2+N)^2
      \left(648-876 N+371 N^2-92 N^3+12 N^4\right)
\nonu \\
&&  +3 k^3
      \left(1536-2249 N+1377 N^2-471 N^3+73 N^4\right)
\nonu \\
&& \left. \left. \left. +k^2
      \left(-7096+13610 N-10495 N^2+4424 N^3-1090 N^4+122
        N^5\right)\right)\right)\right)/
\nonu \\
&& \left(\left(-2+7 N-7 N^2+2
    N^3\right) \left(k^2 (44+5 N)+44 \left(2-3 N+N^2\right)
\right. \right.
\nonu \\
&& \left. \left. +k
    \left(-132+73 N+10 N^2\right)\right)\right),
\nonu \\
c_{13} & = & -\left(3 \left(-c_8 \left(k^2 (44+5 N)+44 \left(2-3
        N+N^2\right)+k \left(-132+73 N+10 N^2\right)\right) \right.
\right.
\nonu \\
&&  +2
    c_{10} \left(44 (-2+N)^3 (-1+N)+6 k^4 (5+2 N)+2 k (-2+N)^2
      \left(-101+73 N+7 N^2\right) \right. \nonu \\
&& \left. \left. \left. +k^3 \left(-238+64 N+41
        N^2\right)+k^2 \left(672-656 N+74 N^2+43
        N^3\right)\right)\right)\right)/
\nonu \\
&&   \left(\left(2-3 N+N^2\right)
  \left(k^2 (44+5 N)+44 \left(2-3 N+N^2\right)+k \left(-132+73 N+10
      N^2\right)\right)\right),
\nonu \\
c_{14} & = & \frac{12 c_{10} (-2+k+N) \left(18+7 k^2-15 N+4 N^2+7
    k (-3+2 N)\right)}{k^2 (44+5 N)+44 \left(2-3 N+N^2\right)+k
  \left(-132+73 N+10 N^2\right)},
\nonu \\
c_{15} & = & -\frac{3 c_{10} \left(18+7 k^2-15 N+4 N^2+7 k (-3+2
    N)\right)}{k^2 (44+5 N)+44 \left(2-3 N+N^2\right)+k \left(-132+73
    N+10 N^2\right)},
\nonu \\
c_{18} & = & \frac{-6 c_8+4 c_{10} \left(3 k^2+5 k (-2+N)+2 (-2+N)^2\right)}{-2+N}, 
\nonu \\
c_{21} & = & \left(k \left(c_8 (-1+k) (-8+N)+2 c_{10}
    \left(6 k^3+2 (-4+N) (-2+N)^2 \right. \right. \right.
\nonu \\
&& \left. \left. \left. +k^2 (-32+13 N)+k \left(56-46 N+9
        N^2\right)\right)\right)\right)/\left(2 \left(-2+7 N-7 N^2+2
    N^3\right)\right),
\nonu \\
c_{22} & = & -\frac{1}{2-5 N+2 N^2}2 \left(c_8 (-1+k) (-8+N) \right. \nonu \\
&& \left. +2
  c_{10} \left(6 k^3+2 (-4+N) (-2+N)^2+k^2 (-32+13 N)+k
    \left(56-46 N+9 N^2\right)\right)\right),
\nonu \\
d_1 & = & \left(k \left(-c_8 (-1+k) (-2+N) \left(k^2 (44+5 N)+44
      \left(2-3 N+N^2\right) \right. \right. \right.
\nonu \\
&&   \left. +k \left(-132+73 N+10 N^2\right)\right)+2
    c_{10} \left(6 k^5 (-8+N)+k^4 \left(472-322 N+25 N^2\right)
\right.
\nonu \\
&&  -8
      (-2+N)^2 \left(-28+55 N-32 N^2+5 N^3\right)+k^3 \left(-1768+2172
        N-705 N^2+35 N^3\right)\nonu \\
&& 
+k^2 \left(3168-5644 N+3294 N^2-663
        N^3+20 N^4\right) \nonu \\
&&  \left. \left. \left. +k \left(-2720+6408 N-5460 N^2+2000 N^3-274
        N^4+4 N^5\right)\right)\right)\right)/
\nonu \\
&&
\left(4 (-1+N) \left(k^2
    (44+5 N)+44 \left(2-3 N+N^2\right)+k \left(-132+73 N+10
      N^2\right)\right)\right),
\nonu \\
d_2 & = & \left(k \left(c_8 (-1+k) \left(-4+N^2\right) \left(k^2
      (44+5 N)+44 \left(2-3 N+N^2\right) \right. \right. \right. \nonu \\
&& \left.  +k \left(-132+73 N+10
        N^2\right)\right)-2 c_{10} \left(18 k^5
      \left(-8+N^2\right) \right. \nonu \\
&& 
+3 k^4 \left(384-152 N-82 N^2+25 N^3\right) \nonu \\
&&  -8
      (-2+N)^2 \left(-40+58 N-8 N^2-13 N^3+3 N^4\right)
\nonu \\
&&  +k^3
      \left(-3632+2992 N+464 N^2-727 N^3+105 N^4\right)
\nonu \\
&&  +k^2
      \left(5632-7304 N+1064 N^2+1778 N^3-705 N^4+60 N^5\right)\nonu 
\\
&& \left. \left. \left.  +2 k
      \left(-2144+3928 N-1764 N^2-552 N^3+590 N^4-127 N^5+6
        N^6\right)\right)\right)\right)/
\nonu \\
&& \left(4 \left(1-3 N+2
    N^2\right) \left(k^2 (44+5 N)+44 \left(2-3 N+N^2\right)+k
    \left(-132+73 N+10 N^2\right)\right)\right), 
\nonu \\
d_3 & = & \left(3 c_{10} \left(2 k^4 (-8+N)+k^3 \left(96-76 N+8
      N^2\right)+k^2 \left(-104+216 N-103 N^2+12 N^3\right)
  \right. \right.
\nonu \\
&& \left. \left. +4
    \left(36-84 N+71 N^2-27 N^3+4 N^4\right)+k \left(-120+62 N+57
      N^2-42 N^3+8 N^4\right)\right)\right)/
\nonu \\
&& \left(2 \left(k^2 (44+5
    N)+44 \left(2-3 N+N^2\right)+k \left(-132+73 N+10
      N^2\right)\right)\right),
\nonu \\
d_4 & = & -\left(c_{10} \left(2 k^4 (-8+N)+k^3 \left(96-76 N+8
      N^2\right)+k^2 \left(-104+216 N-103 N^2+12 N^3\right)
  \right. \right.
\nonu \\
&& \left. \left. +4
    \left(36-84 N+71 N^2-27 N^3+4 N^4\right)+k \left(-120+62 N+57
      N^2-42 N^3+8 N^4\right)\right)\right)/
\nonu \\
&& \left(k^2 (44+5 N)+44
  \left(2-3 N+N^2\right)+k \left(-132+73 N+10 N^2\right)\right), 
\nonu \\
d_5 & = & \left(-6 c_8 (-1+k) (-3+N) N \left(k^2 (44+5 N)+44
    \left(2-3 N+N^2\right) \right. \right.\nonu \\
&&  \left. +k \left(-132+73 N+10
      N^2\right)\right)+c_{10} \left(6 k^5 N \left(-197+15 N+14
      N^2\right) \right.
\nonu \\
&& 
+k^4 \left(96+9502 N-5303 N^2-445 N^3+350 N^4\right)
\nonu \\
&&  +4
    (-2+N)^2 \left(-64+1060 N-1945 N^2+1237 N^3-316 N^4+28
      N^5\right)
\nonu \\
&&  +k^3 \left(-320-31012 N+33947 N^2-8004 N^3-1485
      N^4+490 N^5\right) \nonu \\
&& +k^2 \left(-128+52536 N-87576 N^2+47653
      N^3-7523 N^4-1042 N^5+280 N^6\right) \nonu \\
&& \left. \left. +2 k \left(640-23560 N+52290
      N^2-44415 N^3+17202 N^4-2685 N^5-4 N^6+28
      N^7\right)\right)\right)/
\nonu \\
&& \left(\left(1-3 N+2 N^2\right)
  \left(k^2 (44+5 N)+44 \left(2-3 N+N^2\right)+k \left(-132+73 N+10
      N^2\right)\right)\right),
\nonu \\
d_6 & = & -\left(5 c_8 (-2+N) \left(k^2 (44+5 N)+44 \left(2-3
      N+N^2\right)+k \left(-132+73 N+10 N^2\right)\right)
\right.
\nonu \\
&&  +4 c_{10}
  \left(3 k^5 (-8+N)+k^4 \left(368-212 N+5 N^2\right)-k^3
    \left(1720-1893 N+492 N^2+10 N^3\right) \right. \nonu \\
&& -4 (-2+N)^2 \left(-72+143
      N-87 N^2+16 N^3\right) \nonu \\
&&  +k^2 \left(3520-5888 N+3141 N^2-494 N^3-20
      N^4\right) \nonu \\
&&  \left. \left. -2 k \left(1648-3786 N+3149 N^2-1110 N^3+130 N^4+4
      N^5\right)\right)\right)/
\nonu \\
&& \left(2 \left(k^2 (44+5 N)+44 \left(2-3
      N+N^2\right)+k \left(-132+73 N+10 N^2\right)\right)\right),
\nonu \\
d_7 & = & \left(c_{10} (-2+k+N) \left(2 k^4 (-8+N)+k^3
    \left(176-126 N+13 N^2\right)  \right. \right. \nonu \\
&&  +k^2 \left(-416+584 N-249 N^2+32
      N^3\right)+4 \left(-48-40 N+130 N^2-77 N^3+14 N^4\right)
\nonu \\
&& \left. \left. +2 k
    \left(176-240 N+161 N^2-69 N^3+14
      N^4\right)\right)\right)/
\nonu \\
&& \left(k^2 (44+5 N)+44 \left(2-3
    N+N^2\right)+k \left(-132+73 N+10 N^2\right)\right), 
\nonu \\
d_8 & = & \left(2 \left(-c_8 (-1+k) \left(-10-21 N+7 N^2\right)
    \left(k^2 (44+5 N)+44 \left(2-3 N+N^2\right)
    \right. \right. \right. \nonu \\
&&  \left. \left. +k \left(-132+73 N+10
        N^2\right)\right)+4 c_{10} \left(6 k^5 \left(-28-62
        N+N^2+5 N^3\right) \right. \right.
\nonu \\
&&  +2 (-2+N)^3 \left(-76-233 N+471 N^2-182
        N^3+20 N^4\right)
\nonu \\
&&  +k^4 \left(1436+2274 N-1474 N^2-261 N^3+125
        N^4\right) \nonu \\
&&
+k (-2+N)^2 \left(-1252-2285 N+4489 N^2-1538 N^3+62
        N^4+20 N^5\right) \nonu \\
&&  
+k^3 \left(-4676-5144 N+9067 N^2-1946 N^3-668
        N^4+175 N^5\right)
\nonu \\
&& \left. \left. \left. 
+k^2 \left(7152+5758 N-22139 N^2+13624
        N^3-1946 N^4-449 N^5+100
        N^6\right)\right)\right)\right)/
\nonu \\
&& \left(\left(-2+7 N-7 N^2+2
    N^3\right) \left(k^2 (44+5 N)+44 \left(2-3 N+N^2\right)
\right. \right.
\nonu \\
&& \left. \left. +k
    \left(-132+73 N+10 N^2\right)\right)\right),
\nonu \\
d_9 & = & \left(4 c_{10} (-2+k+N) \left(5 k^3 (-8+N)+2 k^2
    \left(78-53 N+10 N^2\right) \right. \right. \nonu \\
&& \left. \left. +4 \left(42-N-30 N^2+10 N^3\right)+k
    \left(-236+153 N-56 N^2+20 N^3\right)\right)\right)/
\nonu \\
&& \left(k^2
  (44+5 N)+44 \left(2-3 N+N^2\right)+k \left(-132+73 N+10
    N^2\right)\right).
\label{coeffwd}
\eea
The coefficients $c_8$ and $c_{10}$ for $N=8$ are determined by 
(\ref{c8}) and (\ref{c10}). 
We also obtained those coefficients for $N=6$.
For general $N$, they are not known so far.

\section{The coefficients in the spin-$4$ current of
$W B_{\frac{N-1}{2}}$ minimal model }

The explicit coefficient functions \cite{Ahn1202} in (\ref{VB}), in terms of 
$c_9$ and $d_8$, are given by
\bea
c_3  & = &  -\frac{d_8 (-8+N) (-19+7 k+12 N)}{2 (2+k) \left(68-39 
N+10 N^2+k (-4+5 N)\right)}, \nonu \\
c_{10} & = & -\frac{c_9}{4 (-2+k+N)},
\nonu \\
c_{11} & = & -\frac{d_8 k (6 k+11 (-2+N))}
{4 (-1+N) \left(68-39 N+10 N^2+k (-4+5 N)\right)},
\nonu \\
c_{12} & = & \frac{d_8 (6 k+11 (-2+N))}{68-39 N+10 N^2+k (-4+5 N)},
\nonu \\
c_{13} & = & -\frac{3 d_8 (-8+N) \left(2 k^2 (5+2 N)+22 
\left(2-3 N+N^2\right)+k \left(-46+18 N+7 N^2\right)\right)}
{2 (2+k) (-4+2 k+N) \left(2-3 N+N^2\right) \left(68-39 N+10 N^2+k (-4+5 N)\right)},
\nonu \\
c_{14} & = & \left(d_8 (-8+N)^2 (-2+k+N) (-16+11 N) \right. \nonu \\
& + &  
c_9 \left(-2992+4164 N-2456 N^2+779 N^3-129 N^4+10 N^5 \right. \nonu \\
& - &  6 k^3 
\left(4-13 N+10 N^2\right)+k^2 \left(416-1080 N+561 N^2-130
  N^3\right) \nonu \\
& + & \left. \left.
k \left(40+54 N-135 N^2+79 N^3-15 N^4\right)\right)\right)/
\left(\left(68-39 N+10 N^2+k (-4+5 N)\right) \right.  
\nonu \\
& & \left.  \left(-20 (-2+N)^2+6 k^2 
\left(4-N+N^2\right)+k \left(-8+6 N-3 N^2+N^3\right)\right)\right),
\nonu \\
c_{15} & = & 
\left(-d_8 (-8+N)^2 (-2+k+N) (-16+11 N) 
\right. \nonu \\
& + &  c_9 
\left(2992-4164 N+2456 N^2-779 N^3+129 N^4-10 N^5+6 k^3 
\left(4-13 N+10 N^2\right) \right. \nonu \\
& + & \left. \left. k^2 \left(-416+1080 N-561 N^2+130
  N^3\right)
+k \left(-40-54 N+135 N^2-79 N^3+15 N^4\right)\right)\right)/
\nonu \\
& & \left(4 (-2+k+N) \left(68-39 N+10 N^2+k (-4+5 N)\right) \left(-20 
(-2+N)^2 \right. \right. \nonu \\
& + & \left. \left. 6 k^2 \left(4-N+N^2\right)+k \left(-8+6 N-3
  N^2+N^3\right)\right)\right),
\nonu \\
c_{18} & = & \frac{d_8 (-8+N) 
\left(k^2 (44+5 N)+44 \left(2-3 N+N^2\right)+
k \left(-132+73 N+10 N^2\right)\right)}{(2+k) 
(-2+N) (-4+2 k+N) \left(68-39 N+10 N^2+k (-4+5 N)\right)},
\nonu \\
d_1 & = & \frac{3 d_8 k \left(-4+N+2 N^2+k (4+N)\right)}{4 
\left(68-39 N+10 N^2+k (-4+5 N)\right)},
\nonu \\
d_2 & = & -\frac{d_8 k \left(-4+N+2 N^2+k (4+N)\right)}{2 
\left(68-39 N+10 N^2+k (-4+5 N)\right)},
\nonu \\
d_3 & = & -\left(3 \left(-2 d_8 (-8+N)^2 
\left(4 (-2+N)^2 (-1+N)+k^3 \left(4-N+N^2\right)
\right. \right. \right. \nonu \\
&+ & \left.  k^2 
\left(-20+17 N-8 N^2+3 N^3\right)+k \left(32-46 N+25 N^2-9 
N^3+2 N^4\right)\right)\nonu \\
& + & c_9 (-2+N) \left(7616-13072 
N+9648 N^2-3852 N^3+832 N^4-80 N^5 \right. \nonu \\
& + &  2 k^3 \left(48-116 N+82 
N^2-19 N^3+5 N^4\right)
\nonu \\
& + &  k \left(-992-656 N+854 N^2-291 N^3+11 
N^4+10 N^5\right) \nonu \\
& + & \left. \left. \left. 
k^2 \left(-1600+2920 N-1926 N^2+691 N^3-133 N^4+
20 N^5\right)\right)\right)\right)/
\nonu \\
&& \left(8 \left(-136+146 N-59 N^2+
10 N^3+k^2 (-4+5 N)+k \left(76-53 N+15 N^2\right)\right) 
\right. \nonu \\
&& \left. \left(-20 
(-2+N)^2+6 k^2 \left(4-N+N^2\right)+k \left(-8+6 N-3 N^2+N^3\right)
\right)\right),
\nonu \\
d_4 & = & \left(-2 d_8 (-8+N)^2 
\left(4 (-2+N)^2 (-1+N)+k^3 \left(4-N+N^2\right)
\right. \right. \nonu \\
& + &  \left. k^2 
\left(-20+17 N-8 N^2+3 N^3\right)+k \left(32-46 N+25 N^2-9 N^3+2
  N^4\right)\right)
\nonu \\
& + & c_9 (-2+N) \left(7616-13072 N+9648 N^2-3852 N^3+832 N^4-80
  N^5  \right. \nonu \\
& + &  2 
k^3 \left(48-116 N+82 N^2-19 N^3+5 N^4\right)
\nonu \\
& + &  k \left(-992-656 N+854 
N^2-291 N^3+11 N^4+10 N^5\right) \nonu \\
& + & \left. \left. k^2 \left(-1600+2920 N-1926 N^2+691 
N^3-133 N^4+20 N^5\right)\right)\right)/
\nonu \\
&& \left(4 \left(-136+146 N-59 
N^2+10 N^3+k^2 (-4+5 N)+k \left(76-53 N+15 N^2\right)\right) 
\right.
\nonu \\
&& \left. \left(-20 (-2+N)^2+6 k^2 \left(4-N+N^2\right)+k 
\left(-8+6 N-3 N^2+N^3\right)\right)\right),
\nonu \\
d_5 & = & \frac{d_8 \left(-128-8 (-17+k) N+(-61+7 k) N^2+14
    N^3\right)}{4 
\left(68-39 N+10 N^2+k (-4+5 N)\right)},
\nonu \\
d_6 & = & \left(d_8 (-8+N) \left(-k^3 (-8+N)+6 k^2 (-16+9 N)+k 
\left(232-267 N+66 N^2+4 N^3\right)  \right. \right. \nonu \\
& + & \left. \left. 2 \left(-72+143 N-87 N^2+16 N^3
\right)\right)\right)/\left(2 (2+k) (-4+2 k+N) 
\right.
\nonu \\
&& \left. \left(68-39 N+10 
N^2+k (-4+5 N)\right)\right),
\nonu \\
d_7 & = & \left(d_8 (-8+N)^2 
\left(4 (-2+N)^2 (-23+18 N)+2 k^3 \left(4-N+N^2\right)
\right. \right.
\nonu \\
& + & \left. \left. k^2 
\left(-80+64 N-31 N^2+11 N^3\right)+k \left(312-414 N+199 N^2-63 
N^3+14 N^4\right)\right)
\right.
\nonu \\
& - &  4 c_9 (-2+N) \left(7616-13072 N+9648 
N^2-3852 N^3+832 N^4-80 N^5 \right. \nonu \\
& + &  2 k^3 \left(48-116 N+82 N^2-19 N^3+5 
N^4\right)
\nonu \\
& + &  k \left(-992-656 N+854 N^2-291 N^3+11 N^4+10
      N^5\right)
\nonu \\
& + & \left. \left. 
k^2 \left(-1600+2920 N-1926 N^2+691 N^3-133 N^4+20 N^5\right)\right)
\right)/
\nonu \\
&& \left(4 \left(68-39 N+10 N^2+k (-4+5 N)\right) 
\left(-20 (-2+N)^2+6 k^2 \left(4-N+N^2\right)
\right. \right. \nonu \\
& + & \left. \left. k 
\left(-8+6 N-3 N^2+N^3\right)\right)\right),
\nonu \\
d_9 & = & \left(d_8 (-8+N)^2 \left(5 k^2 
\left(4-N+N^2\right)+4 \left(42-53 N+16 N^2\right)
\right. \right. \nonu \\
& + &  \left. k 
\left(-124+99 N-25 N^2+10 N^3\right)\right) \nonu \\
& - &  3 c_9 
\left(7616-13072 N+9648 N^2-3852 N^3+832 N^4-80 N^5
\right. \nonu \\
& + &  2 k^3 
\left(48-116 N+82 N^2-19 N^3+5 N^4\right)\nonu \\
& + &  k \left(-992-656 
N+854 N^2-291 N^3+11 N^4+10 N^5\right) \nonu \\
& + & \left. \left. k^2 \left(-1600+2920 
N-1926 N^2+691 N^3-133 N^4+20
N^5\right)\right)\right)/
\nonu \\
&& \left(\left(68-39 
N+10 N^2+k (-4+5 N)\right) \left(-20 (-2+N)^2+6 k^2
\left(4-N+N^2\right) \right. \right. \nonu \\
&+ & \left. \left.
k \left(-8+6 N-3 N^2+N^3\right)\right)\right).
\label{exactcoeff}
\eea
The coefficients $c_9$ and $d_8$ for $N=7$ are determined by 
(\ref{c9d8}). For general $N$, they are not known so far.
We also obtained those coefficients for $N=5$.



\begin{thebibliography}{99}

\bibitem{GG} 
  M.~R.~Gaberdiel and R.~Gopakumar,
  ``An $AdS_3$ Dual for Minimal Model CFTs,''  
Phys.\ Rev.\ D {\bf 83}, 066007 (2011)  
[arXiv:1011.2986 [hep-th]].  

\bibitem{GG1} 
  M.~R.~Gaberdiel and R.~Gopakumar,
  ``Triality in Minimal Model Holography,''  JHEP {\bf 1207}, 127
  (2012)  
[arXiv:1205.2472 [hep-th]].  

\bibitem{GG2} 
  M.~R.~Gaberdiel and R.~Gopakumar,
  ``Minimal Model Holography,''  arXiv:1207.6697 [hep-th].  

\bibitem{AGKP} 
  M.~Ammon, M.~Gutperle, P.~Kraus and E.~Perlmutter,
  ``Black holes in three dimensional higher spin gravity: A review,''  arXiv:1208.5182 [hep-th].  

\bibitem{BBSS2} 
  F.~A.~Bais, P.~Bouwknegt, M.~Surridge and K.~Schoutens,
  Coset Construction for Extended Virasoro Algebras,''  
Nucl.\ Phys.\ B {\bf 304}, 371 (1988).  

\bibitem{Ahn2011} 
  C.~Ahn,
  ``The Coset Spin-4 Casimir Operator and Its Three-Point Functions
  with Scalars,''  
JHEP {\bf 1202}, 027 (2012)  [arXiv:1111.0091 [hep-th]].  

\bibitem{LF} 
  S.~L.~Lukyanov and V.~A.~Fateev,
  ``Physics reviews: Additional symmetries and 
exactly soluble models in two-dimensional conformal field theory,''  
Chur, Switzerland: Harwood (1990) 117 p. (Soviet Scientific Reviews A, Physics: 15.2)

\bibitem{Ahn1202} 
  C.~Ahn,
  ``The Primary Spin-4 Casimir Operators in the Holographic SO(N)
  Coset Minimal Models,''  
JHEP {\bf 1205}, 040 (2012)  [arXiv:1202.0074 [hep-th]].  

\bibitem{Ahn1106} 
  C.~Ahn,
  ``The Large N 't Hooft Limit of Coset Minimal Models,''  JHEP {\bf
  1110}, 125 (2011)  [arXiv:1106.0351 [hep-th]].  

\bibitem{GV} 
  M.~R.~Gaberdiel and C.~Vollenweider,
  ``Minimal Model Holography for SO(2N),''  JHEP {\bf 1108}, 104 (2011)  [arXiv:1106.2634 [hep-th]].  

\bibitem{Hornfeck} 
  K.~Hornfeck,
  ``Classification of structure constants for W algebras from highest weights,''  
Nucl.\ Phys.\ B {\bf 411}, 307 (1994)  [hep-th/9307170].  

\bibitem{CGKV} 
  C.~Candu, M.~R.~Gaberdiel, M.~Kelm and C.~Vollenweider,
  ``Even spin minimal model holography,''  
arXiv:1211.3113 [hep-th].  

\bibitem{BS} 
  P.~Bouwknegt and K.~Schoutens,
  ``W symmetry in conformal field theory,''  
Phys.\ Rept.\  {\bf 223}, 183 (1993)  [hep-th/9210010].  

\bibitem{GH} 
  M.~R.~Gaberdiel and T.~Hartman,
  ``Symmetries of Holographic Minimal Models,''  JHEP {\bf 1105}, 031 (2011)  [arXiv:1101.2910 [hep-th]].  

\bibitem{GGHR} 
  M.~R.~Gaberdiel, R.~Gopakumar, T.~Hartman and S.~Raju,
  ``Partition Functions of Holographic Minimal Models,''  JHEP {\bf 1108}, 077 (2011)  [arXiv:1106.1897 [hep-th]].  

\bibitem{BBSS1} 
  F.~A.~Bais, P.~Bouwknegt, M.~Surridge and K.~Schoutens,
  ``Extensions of the Virasoro Algebra 
Constructed from Kac-Moody Algebras Using Higher Order Casimir
  Invariants,''  
Nucl.\ Phys.\ B {\bf 304}, 348 (1988).  

\bibitem{Ahn1211} 
  C.~Ahn,
  ``The Higher Spin Currents in the 
N=1 Stringy Coset Minimal Model,''  arXiv:1211.2589 [hep-th].  

\bibitem{Thielemans} 
  K.~Thielemans,
  ``A Mathematica package for 
computing operator product expansions,''  
Int.\ J.\ Mod.\ Phys.\ C {\bf 2}, 787 (1991).  

\bibitem{Watts} 
  G.~M.~T.~Watts,
  ``Wb Algebra Representation Theory,''  Nucl.\ Phys.\ B {\bf 339}, 177 (1990).  

\bibitem{Honecker1} 
  A.~Honecker,
  ``Automorphisms of W algebras and extended rational conformal field theories,''  Nucl.\ Phys.\ B {\bf 400}, 574 (1993)  [hep-th/9211130].  

\bibitem{BEHHH} 
  R.~Blumenhagen, W.~Eholzer, A.~Honecker, K.~Hornfeck and R.~Hubel,
  ``Coset realization of unifying W algebras,''  Int.\ J.\ Mod.\ Phys.\ A {\bf 10}, 2367 (1995)  [hep-th/9406203].  

\bibitem{Bowcock} 
  P.~Bowcock,
  ``Quasi-primary Fields And Associativity Of Chiral Algebras,''  
Nucl.\ Phys.\ B {\bf 356}, 367 (1991).  

\bibitem{BFKNRV}
  R.~Blumenhagen, M.~Flohr, A.~Kliem, W.~Nahm, A.~Recknagel and R.~Varnhagen,
  ``W algebras with two and three generators,''
  Nucl.\ Phys.\  B {\bf 361}, 255 (1991).

\bibitem{Bouwknegt} 
  P.~Bouwknegt,
  ``Extended Conformal Algebras,''  
Phys.\ Lett.\ B {\bf 207}, 295 (1988).  

\bibitem{HT} 
  K.~-J.~Hamada and M.~Takao,
  ``Spin 4 Current Algebra,''  Phys.\ Lett.\ B {\bf 209}, 247 (1988)  [Erratum-ibid.\ B {\bf 213}, 564 (1988)].  

\bibitem{Zhang} 
  D.~-H.~Zhang,
  ``Spin 4 Extended Conformal Algebra,''  Phys.\ Lett.\ B {\bf 232}, 323 (1989).  

\bibitem{KW} 
  H.~G.~Kausch and G.~M.~T.~Watts,
  ``A Study of W algebras using Jacobi identities,''  Nucl.\ Phys.\ B {\bf 354}, 740 (1991).  

\bibitem{Honecker} 
  A.~Honecker,
  ``A Note on the algebraic evaluation of correlators in local chiral conformal field theory,''  hep-th/9209029.  

\bibitem{FRS} 
  L.~Frappat, E.~Ragoucy and P.~Sorba,
  ``Folding the W algebras,''  Nucl.\ Phys.\ B {\bf 404}, 805 (1993)  [hep-th/9301040].  

\bibitem{Hornfeck1992} 
  K.~Hornfeck,
  ``The Minimal supersymmetric extension of WA(n-1),''  Phys.\ Lett.\ B {\bf 275}, 355 (1992).  

\bibitem{LF1} 
  S.~L.~Lukyanov and V.~A.~Fateev,
  ``Exactly Solvable Models Of Conformal Quantum Theory Associated With Simple Lie Algebra D(n). (in Russian),''  
Sov.\ J.\ Nucl.\ Phys.\  {\bf 49}, 925 (1989)  [Yad.\ Fiz.\  {\bf 49}, 1491 (1989)].  

\bibitem{Ahn1206} 
  C.~Ahn,
  ``The Large N 't Hooft Limit of Kazama-Suzuki Model,''  JHEP {\bf 1208}, 047 (2012)  
[arXiv:1206.0054 [hep-th]].  

\bibitem{Ahn1208} 
  C.~Ahn,
  ``The Operator Product Expansion of the Lowest Higher Spin Current at Finite N,''  JHEP {\bf 1301}, 041 (2013)  [arXiv:1208.0058 [hep-th]].  

\bibitem{CLW} 
  B.~Chen, J.~Long and Y.~-n.~Wang,
  ``Black holes in Truncated Higher Spin AdS$_3$ Gravity,''  JHEP {\bf 1212}, 052 (2012)  [arXiv:1209.6185 [hep-th]].  

\bibitem{BW} 
  P.~Bowcock and G.~M.~T.~Watts,
  ``On the classification of quantum W algebras,''  Nucl.\ Phys.\ B {\bf 379}, 63 (1992)  [hep-th/9111062].  

\bibitem{Ahn1992} 
  C.~Ahn,
  ``Explicit construction of spin 4 Casimir operator 
in the coset model SO(5)-1 x SO(5)-m / SO(5)-(1+m),''  J.\ Phys.\ A {\bf 27}, 231 (1994)  [hep-th/9209001].  

\bibitem{Ahn1991} 
  C.~Ahn,
  ``c = 5/2 free fermion model of WB(2) algebra,''  
Int.\ J.\ Mod.\ Phys.\ A {\bf 7}, 6799 (1992)  [hep-th/9111061].  

\bibitem{Ozer} 
  H.~T.~Ozer,
  ``Miura - like free field realization of fermionic Casimir WB(3) algebras,''  
Mod.\ Phys.\ Lett.\ A {\bf 14}, 469 (1999)  [hep-th/9810208].  

\bibitem{FST} 
  J.~M.~Figueroa-O'Farrill, S.~Schrans and K.~Thielemans,
  ``On the Casimir algebra of B(2),''  Phys.\ Lett.\ B {\bf 263}, 
378 (1991).  

\bibitem{CHR} 
  T.~Creutzig, Y.~Hikida and P.~B.~Ronne,
  ``Higher spin $AdS_3$ supergravity and its dual CFT,''  
JHEP {\bf 1202}, 109 (2012)  [arXiv:1111.2139 [hep-th]].  

\bibitem{CG} 
  C.~Candu and M.~R.~Gaberdiel,
  ``Supersymmetric holography on $AdS_3$,''  
arXiv:1203.1939 [hep-th].  

\bibitem{Reyetal} 
  M.~Henneaux, G.~Lucena Gomez, J.~Park and S.~-J.~Rey,
  ``Super- W(infinity) Asymptotic Symmetry of 
Higher-Spin $AdS_3$ Supergravity,''  JHEP {\bf 1206}, 037 (2012)  
[arXiv:1203.5152 [hep-th]].  

\bibitem{HP} 
  K.~Hanaki and C.~Peng,
  ``Symmetries of Holographic Super-Minimal Models,''  
arXiv:1203.5768 [hep-th].  

\bibitem{CG1} 
  C.~Candu and M.~R.~Gaberdiel,
  ``Duality in N=2 minimal model holography,''  
arXiv:1207.6646 [hep-th].  

\bibitem{Tan} 
  H.~S.~Tan,
  ``Exploring Three-dimensional Higher-Spin Supergravity based on sl(N |N - 1) Chern-Simons theories,''  JHEP {\bf 1211}, 063 (2012)  [arXiv:1208.2277 [hep-th]].  

\bibitem{CHR1} 
  T.~Creutzig, Y.~Hikida and P.~B.~Ronne,
  ``N=1 supersymmetric higher spin holography on $AdS_3$,''  arXiv:1209.5404 [hep-th].  

\bibitem{CHR2} 
  T.~Creutzig, Y.~Hikida and P.~B.~Ronne,
  ``Three point functions in higher spin $AdS_3$ supergravity,''  JHEP {\bf 1301}, 171 (2013)  [arXiv:1211.2237 [hep-th]].  

\bibitem{MZ} 
  H.~Moradi and K.~Zoubos,
  ``Three-Point Functions in N=2 Higher-Spin Holography,''  arXiv:1211.2239 [hep-th].  

\bibitem{Peng} 
  C.~Peng,
  ``Dualities from higher-spin supergravity,''  arXiv:1211.6748 [hep-th].  

\bibitem{Hikida} 
  Y.~Hikida,
  ``Conical defects and N=2 higher spin holography,''  arXiv:1212.4124 [hep-th].  

\end{thebibliography}
\end{document}